\documentclass[twocolumn,showpacs,preprintnumbers,amsmath,amssymb,superscriptaddress,aps,prl,longbibliography]{revtex4-1}

\usepackage[pdftex]{graphicx}
\usepackage{xcolor}
\usepackage{amsmath}

\allowdisplaybreaks
\begin{document}
\title{Ultrafast giant enhancement of second harmonic generation \\ through level occupation engineering}

\author{Jun-Yi Shan}
\email[Corresponding author: ]{jyshan0908@gmail.com}
\affiliation{Department of Physics, University of California, Berkeley, California, USA\looseness=-1}
\affiliation{BC Innovative Center, Pennsylvania, USA\looseness=-1}

\date{\today}

\maketitle

\textbf{Optical nonlinearity, especially the second harmonic generation (SHG), is generally weak in materials but has the potential to be applied in high-speed optical computers and energy-efficient artificial intelligence systems. In order to program such photonic circuits, electrical and all-optical modulation mechanisms of optical nonlinearity have been proposed. Among them the electrical methods are bottlenecked by speed, while optical methods like Floquet engineering provides a fast heat-free route, but has only been experimentally shown to suppress SHG. Here we theoretically and experimentally demonstrated an ultrafast enhancement of SHG by 40\% on a timescale of $\sim$ 500 femtosecond in van der Waals NiPS$_3$. We performed single-ion model calculations to show that by optically control the electron occupation of different energy levels, the SHG can be enhanced due to different electronic states involved in the SHG process. We then performed temperature-dependent time-resolved measurements in both linear and nonlinear optics, which confirm our calculations. We also discussed the implications for other materials in the transition metal thiophosphates (MPX$_3$) family.}

\subsection{I. Introduction}

Nonlinear optical processes have been widely used in frequency conversion \cite{cerullo2003ultrafast}, ultrafast pulse generation \cite{haus1975theory}, and characterization \cite{shan2020evidence} and control \cite{shan2024dynamic} of material properties. The application of optical nonlinearity in photonic computing has emerged as a particularly promising frontier, where nonlinear optics is essential for performing computations beyond the linear regime \cite{wright2022deep}, whose prospects were further boosted by on-chip ultrafast lasers \cite{guo2023ultrafast}. In order to program such photonic circuits, electrical tuning methods are bottlenecked by their slow speeds, while optical gating enjoys the fast speeds provided by the short laser pulses and the fast timescale of electron-photon interactions \cite{bogaerts2020programmable}. While all-optical tuning of linear optical properties has been demonstrated extensively, all-optical tuning of nonlinear optical properties, especially the enhancement of optical nonlinearity, is still in the nascent stage. A promising route is the coherent Floquet engineering of the energy levels \cite{shan2021giant}, but only optical nonlinearity suppression, not enhancement, has been reported and the absorption-free conditions required for Floquet mechanisms are hard to be met.

Optical second harmonic generation (SHG) is generally a weak nonlinear process whose strength is set at the synthesis stage. Therefore, optically tuning of SHG in both up and down directions is a particularly useful first step towards on-demand optical nonlinearity modulation. Noncentrosymmetric systems generally exhibit large SHG, and therefore, a common observation in time-resolved SHG measurements is a suppressed SHG when the laser induces a inversion-symmetry-restoring phase transition \cite{sie2019ultrafast,de2022decoupling,shan2024dynamic}. However, there has been few, if any, reports of an inversion-symmetry-breaking light-induced phase transition which can lead to an enhanced SHG.

In order to solve this obstacle SHG enhancement, we start from understanding the microscopic mechanism of SHG. As a second-order process, SHG involves three states, the initial, intermediate, and final states, which we denote as $\vert i\rangle$, $\vert m\rangle$, and $\vert f\rangle$ (Fig. 1a). In noncentrosymmetric systems, the SHG response is dominated by a bulk electric-dipole (ED) process of the form $P_i(2\omega)=\chi_{ijk}^\mathrm{ED}E_{j}^\mathrm{pr}(\omega)E_{k}^\mathrm{pr}(\omega)$, where the second-order susceptibility tensor $\chi_{ijk}^\mathrm{ED}$ governs the relationship between the incident (probe) electric field $E_{i}^\mathrm{pr}(\omega)$ and the polarization induced at twice the incident probing frequency $P_i(2\omega)$, and the indices $i$, $j$, $k$ run over the \textit{x}, \textit{y}, and \textit{z} coordinates \cite{boyd,shan2021giant}. The quantum mechanical expression for $\chi_{ijk}^\mathrm{ED}$ is given by

\begin{equation}\label{eqn:1}
\begin{split}
    \chi_{ijk}^\mathrm{ED}\propto\sum&\frac{\langle i\vert r_i\vert f\rangle\langle f\vert r_j\vert m\rangle\langle m\vert r_k\vert i\rangle}{(E_f-E_i-2\hbar\omega)(E_m-E_i-\hbar\omega)}\\&+(j\leftrightarrow k)
    \end{split}
\end{equation}

\noindent where the sum is performed over all relevant ions. 

\begin{figure}[t]
\includegraphics[width=0.8\columnwidth]{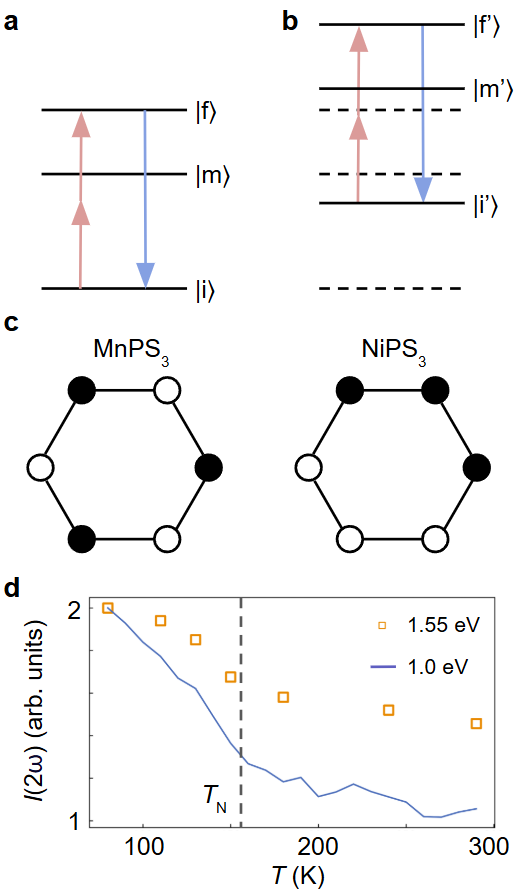}
\caption{\label{fig1} \textbf{Magnetism and SHG in MPX$_3$. a.} Illustration of the SHG process. Red arrows indicate the fundamental photon energy and the blue arrow indicates the SHG photon energy. \textbf{b,} When the electronic level occupation is engineered, a new set of states are involved in the SHG process. \textbf{c,} The magnetic configurations of transition metal ions in MnPS$_3$ and NiPS$_3$. White and black circles indicate opposite spin directions. \textbf{d,} The temperature dependence of the static SHG in NiPS$_3$, for both $\hbar\omega$ = 1.55 eV and 1.0 eV.}
\end{figure}

We optically enhance the SHG through engineering the electronic level occupation, so the SHG process involves a new set of electronic states other than $\vert i\rangle$, $\vert m\rangle$, and $\vert f\rangle$ (Fig. 1b). Note that $\vert i\rangle$ is the ground state of the system, and $\vert m\rangle$ and $\vert f\rangle$ are the states with energies closest to $\hbar\omega$ and $2\hbar\omega$. If many electrons are pre-excited to another level $\vert i'\rangle$, the corresponding intermediate and final states will become $\vert m'\rangle$ and $\vert f'\rangle$ for the SHG processes involving these electrons, and the SHG intensity will be modulated. To contrast with the Floquet engineering mechanism, the Floquet approach focused on optically modulating the energy of the same electronic states, $E_i$, $E_m$, and $E_f$, while our level occupation engineering approach uses a new set of electronic states altogether.

We chose to perform calculations and experiments on MPX$_3$ (M = Mn, Ni, Fe; X = S, Se) materials, especially NiPS$_3$, for several reasons. First of all, these van der Waals materials exhibit rich physics \cite{chu2020linear,kang2020coherent,chen2021electrically,shan2021giant,belvin2021exciton,he2024magnetically}. In particular, while they all exhibit AFM order below their respective Néel temperatures $T_\mathrm{N}$, MnPS$_3$ exhibits an inversion-symmetry-breaking Néel AFM order, FePS$_3$ exhibits an inversion-symmetry-preserving out-of-plane zigzag AFM order, and NiPS$_3$ exhibits an inversion-symmetry-preserving in-plane zigzag AFM order \cite{chu2020linear}. Therefore, MnPS$_3$ enables ED-SHG below $T_\mathrm{N}$ while in the other two ED-SHG is forbidden (Fig. 1c). We used NiPS$_3$ in our time-resolved measurements to isolate the charge-induced effects which we are concerned about, and to minimize the magnetism-induced effects on SHG. Another reason why we chose this material family is that their SHG process can be well approximated by a single-ion picture \cite{shan2021giant,kim2018charge}. As we show below, the electronic states at relevant energy scales are $d$ electrons from the transition metal ions and $p$ electrons from ligand ions, making the level engineering viable and the calculations feasible.

\subsection{II. Results}

To measure SHG in NiPS$_3$ we adopted the fast-rotating CCD-based SHG rotational anisotropy setup \cite{harter2015high}. We fixed the scattering plane angle at the maximum SHG intensity, recorded the SHG intensity $I(2\omega)$ with both the incident and reflected polarizations perpendicular to the scattering plane, and measured its temperature dependence \cite{shan2020evidence}. Contrary to prior reports \cite{chu2020linear}, we observed higher SHG at lower temperatures for two different fundamental photon energies $\hbar\omega$ = 1.55 eV and 1.0 eV (Fig. 1d). The higher SHG could be caused by magneto-elastic coupling induced by magnetic order or higher magnetic correlations at lower temperatures, and has an electric-quadrupole nature \cite{boyd}. We note that this discrepancy from previous works might be a result of measuring bulk NiPS$_3$ or films of different thicknesses \cite{lane2020thickness}. The SHG increase we observed in NiPS$_3$ is much less significant than in MnPS$_3$, where the AFM order breaks inversion symmetry \cite{shan2021giant, chu2020linear}.

We then theoretically study the effects of an engineered level occupation on the SHG intensity of a fixed fundamental probe photon energy of $\hbar\omega=1.55$ eV. First, we note that since the SHG process in NiPS$_3$ is of the electric-quadrupole type of $P^\mathrm{EQ}_i(2\omega)=\chi_{ijkl}^\mathrm{EQ}E_{j}^\mathrm{pr}(\omega)\nabla_kE_{l}^\mathrm{pr}(\omega)$, Eq. (1) should be modified so that one of the electric-dipole matrix elements $\langle\phi_1\vert\mathbf{r}\vert\phi_2\rangle$ becomes an electric-quadrupole matrix element $\mathbf{k}\cdot\langle\phi_1\vert\mathbf{r}\mathbf{r}\vert\phi_2\rangle$ where $\mathbf{k}$ is the light wave vector. Second, we figure out that in the static case, the $\vert i\rangle\rightarrow\vert m\rangle$ transition at 1.55 eV is mainly a $d$-$d$ transition while the $\vert i\rangle\rightarrow\vert f\rangle$ transition is a charge-transfer transition (Fig. 2c to 2e) \cite{kim2018charge}. The electronic ground state of the system has a $d^9\underline{L}$ character where $\underline{L}$ indicates a sulfur hole, while the nine Ni $3d$ electrons have a $t_{2g}^6e_g^3$ configuration with $^3A_{2g}$ symmetry. The charge transfer transition occurs between Ni and S ions on neighboring NiS$_6$ clusters.

\begin{figure*}[t]
\includegraphics[width=2\columnwidth]{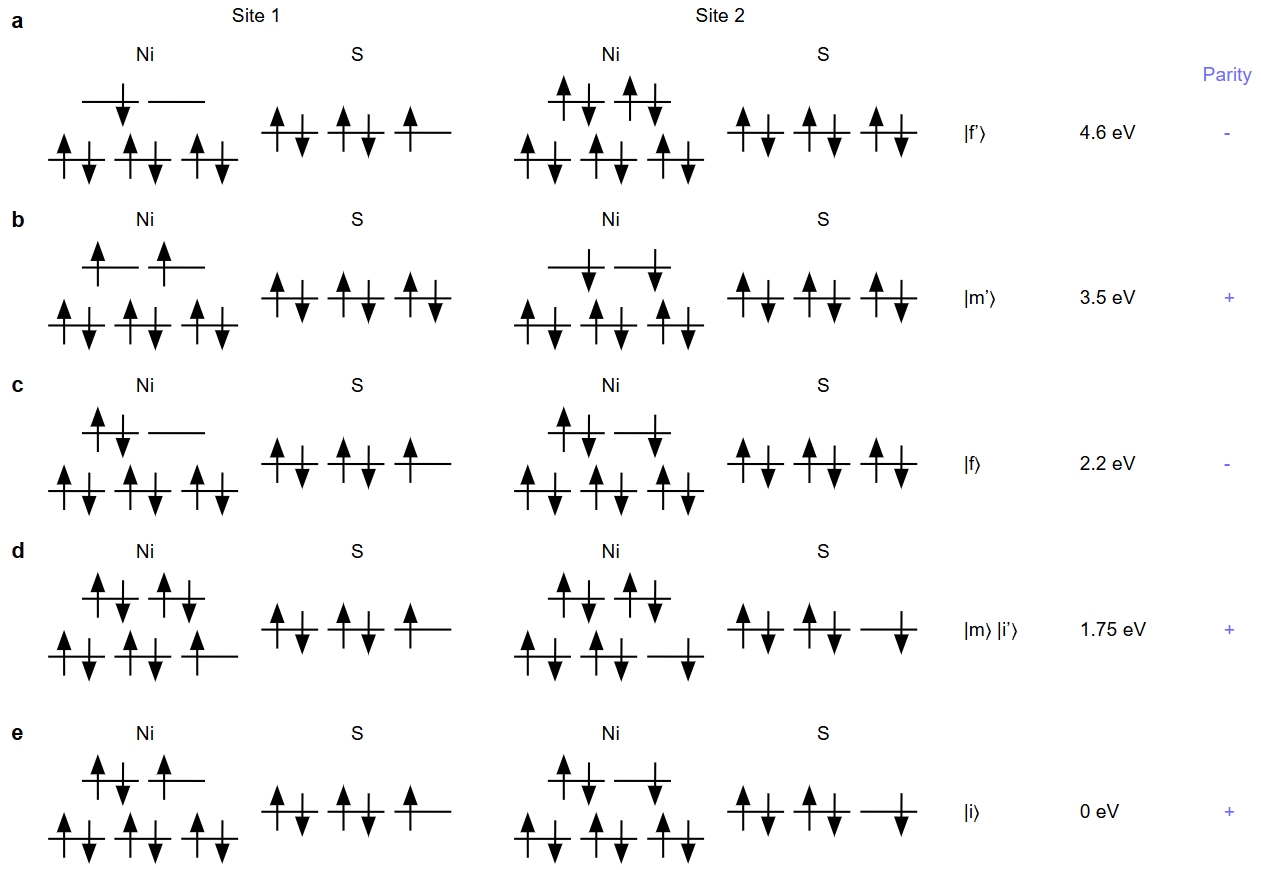}
\caption{\label{fig2} \textbf{Electronic configurations of states involved in the static and engineered SHG processes.} We show the electronic configuration on Ni and S ions for two neighboring NiS$_6$ clusters. On the right we label their roles in the static and engineered SHG processes as well as their respective energies and parities.}
\end{figure*}

If we engineer the level occupation by pumping the ground state of NiPS$_3$ with $\hbar\omega_\mathrm{pu}=1.90$ eV photons, a quasi-static population at an excited level $\vert i'\rangle$ would be accumulated. In our case, $\vert i'\rangle$ coincides with $\vert m\rangle$ because it is the closest level below $\hbar\omega_\mathrm{pu}$. The $\vert i\rangle\rightarrow\vert i'\rangle$ transition is dipole allowed because the trigonal crystal field mixes parities. As a result, the $\vert m'\rangle$ and $\vert f'\rangle$ states involved in the new SHG process would have even higher energies. In optical spectrum there are two strong peaks at $\sim3.5$ and 4.6 eV, making them good candidates for $\vert m'\rangle$ and $\vert f'\rangle$. We do not know their exact electronic configurations, but we postulate that $\vert m'\rangle$ corresponds to an onsite charge transfer between Ni and S from $\vert i\rangle$, and $\vert f'\rangle$ corresponds to a two-electron charge transfer between Ni and S ions on neighboring NiS$_6$ clusters from $\vert i\rangle$ (Fig. 2a and 2b). It is straightforward to verify that both ($\vert i\rangle$, $\vert m\rangle$, $\vert f\rangle$) and ($\vert i'\rangle$, $\vert m'\rangle$, $\vert f'\rangle$) are allowed in electric-quadrupole SHG processes by parity and spin.

\begin{figure}[]
\includegraphics[width=0.8\columnwidth]{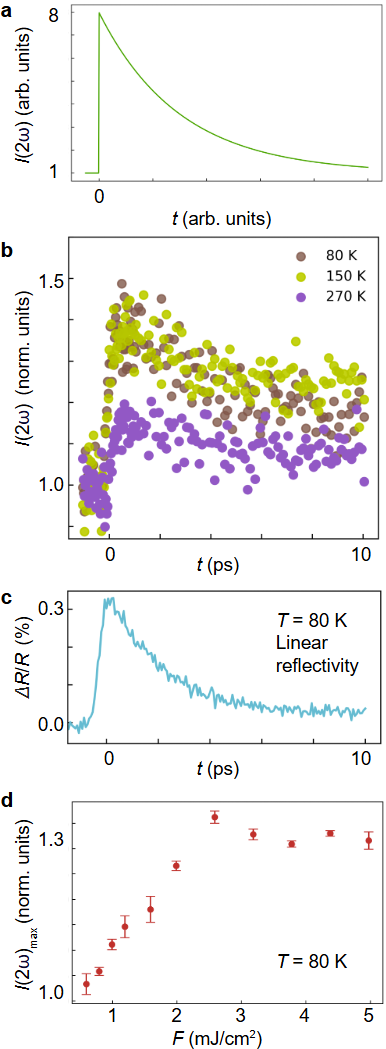}
\caption{\label{fig3} \textbf{Giant enhancement of SHG by level occupation engineering. a.} A simple model estimating the ultrafast enhancement and subsequent decay of SHG intensity if 50\% ground state electrons are promoted to $\vert i'\rangle$. \textbf{b,} The measured time-resolved SHG with fluence $F=2.6$ mJ/cm$^2$ at three different temperatures. Each curve is normalized to its $t<0$ value. \textbf{c,} The measured transient linear reflectivity $\Delta R/R$ at 80 K and $F=2.6$ mJ/cm$^2$. \textbf{d,} The measured maximum post-pump 80 K SHG intensity $I(2\omega, t=0.5$ ps) as a function of fluence $F$. The error bars represent the standard error of the mean from three independent measurements.}
\end{figure}

To compute the SHG intensities in both the static and engineered cases, we need the matrix elements and the energy detuning [see Eq. (1) although one matrix element should be replaced by an electric-quadrupole one as formulated previously]. Since the matrix elements are all allowed without further need of hybridization or spin-orbit coupling \cite{shan2021giant}, we can safely assume that the matrix element component of SHG is similar between the static and engineered cases. Therefore, the SHG discrepancy between the static and engineered cases are dominated by the differences in the detuning, i.e., $1/[(E_f-E_i-2\hbar\omega)(E_m-E_i-\hbar\omega)]$ and $1/[(E_{f'}-E_{i'}-2\hbar\omega)(E_{m'}-E_{i'}-\hbar\omega)]$, which means a $12.96\times$ SHG intensity enhancement if the SHG process starts from $\vert i'\rangle$. If we further assume that after the pump pulse arrives, 50\% of electrons are promoted from $\vert i\rangle$ to $\vert i'\rangle$, and the total measured SHG intensity is an incoherent sum between SHG starting from $\vert i\rangle$ and $\vert i'\rangle$, we expect to see an eight-fold increase of SHG intensity immediately after the pump pulse arrives, and an exponential decay due to the cooling of the electron temperature (Fig. 3a).

In reality, due to the weakness of the trigonal crystal field and a rapid electronic cooling process within the pump pulse \cite{kirilyuk2010ultrafast}, much fewer electrons are promoted at the end of the pump pulse. We set out to experimentally test our proposal and assess the effectiveness of the level occupation engineering. We carried out time-resolved SHG experiments on bulk NiPS$_3$ with a fundamental probe photon energy $\hbar\omega=1.55$ eV and a linearly-polarized pump photon energy of $\hbar\omega_\mathrm{pu}=1.90$ eV. The fundamental output of a Ti:sapphire femtosecond laser at 800 nm was used as the probe pulse (80 fs width), which was focused obliquely onto a 60-$\mu$m spot on the cleaved surface of the NiPS$_3$ crystal at a 10$^\circ$ angle of incidence with a fluence of 1.4 mJ/cm$^{-2}$. The pump photon is generated through an SHG process by 0.95 eV photons from an optical parametric amplifier (OPA). The 0.95 eV photon transmits through a thick $\beta$-BaB$_2$O$_4$ (BBO) crystal and is focused normally onto an 80-$\mu$m spot on the sample.   

As expected, the time-resolved SHG results show an ultrafast giant enhancement of $\sim40$\% after the 1.90 eV pump pulse arrives, following an exponential decay (Fig. 3b). The SHG enhancement persists at various temperatures, including below $T_\mathrm{N}$, close to $T_\mathrm{N}$, and above $T_\mathrm{N}$, demonstrating that the SHG enhancement has negligible correlation with the AFM order parameter. The electronic origin of the SHG enhancement can be further corroborated by the similarity between the SHG transient and linear reflectivity transient $\Delta R/R$ (Fig. 3c). The slow electronic cooling might be attributed to the metastability of the $\vert i'\rangle$ state \cite{ho2021band,ergeccen2022magnetically}. The SHG enhancement saturates at $F\sim2.6$ mJ/cm$^2$ (Fig. 3d), which corresponds to a 3.3\% $\vert i\rangle\rightarrow\vert i'\rangle$ electron population promotion by the pump pulse. We did not observe any pump polarization dependence of the SHG enhancement.

\subsection{III. Discussions}

We anticipate that the SHG enhancement can be observed in all compounds in the MPX$_3$ family, though the signal can be complicated by other features. In MnPS$_3$ for example, since SHG is mainly induced by the noncentrosymmetric AFM order, we should take into account the fact that an ultrafast laser pulse can also melt the AFM order if the pump photon energy is above the $d$-$d$ transition gap, or if the subgap laser pulse is intense enough and long enough, as in commercial picosecond lasers. Therefore, we expect that close to its Néel temperature $T_\mathrm{N}=78$ K, the pump light would mainly suppress the SHG due to the carrier excitation and collapse of the AFM order, while at lower temperatures, the SHG enhancement engineered by level occupation is more dominant. Even for low-energy pump photons, level occupation engineering can be enabled by multi-photon absorption with intense and long pulses.

It is beneficial to expand the suite of mechanisms how light can modulate the optical properties of materials, linear or nonlinear, enhancement or suppression. Our work here and previous works \cite{shan2021giant} show that the MPX$_3$ materials are a versatile platform and pave the way for high-speed optical computers that can perform complex nonlinear tasks.

\section*{A\lowercase{cknowledgements}}

J.-Y.S. collected the experimental data during his Ph.D. studies at Principal Investigator David Hsieh's laboratory at the California Institute of Technology.

\section*{A\lowercase{uthor contributions}}
J.-Y.S. conceived the project, performed the theoretical analyses and the optical experiments, and wrote the manuscript.

\section*{C\lowercase{ompeting} I\lowercase{nterests}}
Authors declare that they have no competing interests.

\section*{D\lowercase{ata} A\lowercase{vailability}}
The data that support the findings of this study are available from the corresponding author upon request.

\bibliography{citations.bib}

\begin{thebibliography}{22}%
\makeatletter
\providecommand \@ifxundefined [1]{%
 \@ifx{#1\undefined}
}%
\providecommand \@ifnum [1]{%
 \ifnum #1\expandafter \@firstoftwo
 \else \expandafter \@secondoftwo
 \fi
}%
\providecommand \@ifx [1]{%
 \ifx #1\expandafter \@firstoftwo
 \else \expandafter \@secondoftwo
 \fi
}%
\providecommand \natexlab [1]{#1}%
\providecommand \enquote  [1]{``#1''}%
\providecommand \bibnamefont  [1]{#1}%
\providecommand \bibfnamefont [1]{#1}%
\providecommand \citenamefont [1]{#1}%
\providecommand \href@noop [0]{\@secondoftwo}%
\providecommand \href [0]{\begingroup \@sanitize@url \@href}%
\providecommand \@href[1]{\@@startlink{#1}\@@href}%
\providecommand \@@href[1]{\endgroup#1\@@endlink}%
\providecommand \@sanitize@url [0]{\catcode `\\12\catcode `\$12\catcode `\&12\catcode `\#12\catcode `\^12\catcode `\_12\catcode `\%12\relax}%
\providecommand \@@startlink[1]{}%
\providecommand \@@endlink[0]{}%
\providecommand \url  [0]{\begingroup\@sanitize@url \@url }%
\providecommand \@url [1]{\endgroup\@href {#1}{\urlprefix }}%
\providecommand \urlprefix  [0]{URL }%
\providecommand \Eprint [0]{\href }%
\providecommand \doibase [0]{http://dx.doi.org/}%
\providecommand \selectlanguage [0]{\@gobble}%
\providecommand \bibinfo  [0]{\@secondoftwo}%
\providecommand \bibfield  [0]{\@secondoftwo}%
\providecommand \translation [1]{[#1]}%
\providecommand \BibitemOpen [0]{}%
\providecommand \bibitemStop [0]{}%
\providecommand \bibitemNoStop [0]{.\EOS\space}%
\providecommand \EOS [0]{\spacefactor3000\relax}%
\providecommand \BibitemShut  [1]{\csname bibitem#1\endcsname}%
\let\auto@bib@innerbib\@empty
\bibitem [{\citenamefont {Cerullo}\ and\ \citenamefont {De~Silvestri}(2003)}]{cerullo2003ultrafast}%
  \BibitemOpen
  \bibfield  {author} {\bibinfo {author} {\bibfnamefont {Giulio}\ \bibnamefont {Cerullo}}\ and\ \bibinfo {author} {\bibfnamefont {Sandro}\ \bibnamefont {De~Silvestri}},\ }\bibfield  {title} {\enquote {\bibinfo {title} {Ultrafast optical parametric amplifiers},}\ }\href@noop {} {\bibfield  {journal} {\bibinfo  {journal} {Rev. Sci. Instrum.}\ }\textbf {\bibinfo {volume} {74}},\ \bibinfo {pages} {1--18} (\bibinfo {year} {2003})}\BibitemShut {NoStop}%
\bibitem [{\citenamefont {Haus}(1975)}]{haus1975theory}%
  \BibitemOpen
  \bibfield  {author} {\bibinfo {author} {\bibfnamefont {Hermann}\ \bibnamefont {Haus}},\ }\bibfield  {title} {\enquote {\bibinfo {title} {Theory of mode locking with a slow saturable absorber},}\ }\href@noop {} {\bibfield  {journal} {\bibinfo  {journal} {IEEE J. Quantum Electron.}\ }\textbf {\bibinfo {volume} {11}},\ \bibinfo {pages} {736--746} (\bibinfo {year} {1975})}\BibitemShut {NoStop}%
\bibitem [{\citenamefont {Shan}\ \emph {et~al.}(2020)\citenamefont {Shan}, \citenamefont {De~la Torre}, \citenamefont {Laurita}, \citenamefont {Zhao}, \citenamefont {Dashwood}, \citenamefont {Puggioni}, \citenamefont {Wang}, \citenamefont {Yamaura}, \citenamefont {Shi}, \citenamefont {Rondinelli} \emph {et~al.}}]{shan2020evidence}%
  \BibitemOpen
  \bibfield  {author} {\bibinfo {author} {\bibfnamefont {Jun-Yi}\ \bibnamefont {Shan}}, \bibinfo {author} {\bibfnamefont {A}~\bibnamefont {De~la Torre}}, \bibinfo {author} {\bibfnamefont {NJ}~\bibnamefont {Laurita}}, \bibinfo {author} {\bibfnamefont {L}~\bibnamefont {Zhao}}, \bibinfo {author} {\bibfnamefont {CD}~\bibnamefont {Dashwood}}, \bibinfo {author} {\bibfnamefont {D}~\bibnamefont {Puggioni}}, \bibinfo {author} {\bibfnamefont {CX}~\bibnamefont {Wang}}, \bibinfo {author} {\bibfnamefont {K}~\bibnamefont {Yamaura}}, \bibinfo {author} {\bibfnamefont {Y}~\bibnamefont {Shi}}, \bibinfo {author} {\bibfnamefont {JM}~\bibnamefont {Rondinelli}},  \emph {et~al.},\ }\bibfield  {title} {\enquote {\bibinfo {title} {Evidence for an extended critical fluctuation region above the polar ordering transition in {LiOsO$_3$}},}\ }\href@noop {} {\bibfield  {journal} {\bibinfo  {journal} {Phys. Rev. Res.}\ }\textbf {\bibinfo {volume} {2}},\ \bibinfo {pages} {033174} (\bibinfo {year} {2020})}\BibitemShut {NoStop}%
\bibitem [{\citenamefont {Shan}\ \emph {et~al.}(2024)\citenamefont {Shan}, \citenamefont {Curtis}, \citenamefont {Guo}, \citenamefont {Roh}, \citenamefont {Rotundu}, \citenamefont {Lee}, \citenamefont {Narang}, \citenamefont {Noh}, \citenamefont {Demler},\ and\ \citenamefont {Hsieh}}]{shan2024dynamic}%
  \BibitemOpen
  \bibfield  {author} {\bibinfo {author} {\bibfnamefont {Jun-Yi}\ \bibnamefont {Shan}}, \bibinfo {author} {\bibfnamefont {Jonathan~B}\ \bibnamefont {Curtis}}, \bibinfo {author} {\bibfnamefont {Mingyao}\ \bibnamefont {Guo}}, \bibinfo {author} {\bibfnamefont {Chang~Jae}\ \bibnamefont {Roh}}, \bibinfo {author} {\bibfnamefont {CR}~\bibnamefont {Rotundu}}, \bibinfo {author} {\bibfnamefont {Young~S}\ \bibnamefont {Lee}}, \bibinfo {author} {\bibfnamefont {Prineha}\ \bibnamefont {Narang}}, \bibinfo {author} {\bibfnamefont {Tae~Won}\ \bibnamefont {Noh}}, \bibinfo {author} {\bibfnamefont {Eugene}\ \bibnamefont {Demler}}, \ and\ \bibinfo {author} {\bibfnamefont {D}~\bibnamefont {Hsieh}},\ }\bibfield  {title} {\enquote {\bibinfo {title} {Dynamic magnetic phase transition induced by parametric magnon pumping},}\ }\href@noop {} {\bibfield  {journal} {\bibinfo  {journal} {Phys. Rev. B}\ }\textbf {\bibinfo {volume} {109}},\ \bibinfo {pages} {054302} (\bibinfo {year} {2024})}\BibitemShut {NoStop}%
\bibitem [{\citenamefont {Wright}\ \emph {et~al.}(2022)\citenamefont {Wright}, \citenamefont {Onodera}, \citenamefont {Stein}, \citenamefont {Wang}, \citenamefont {Schachter}, \citenamefont {Hu},\ and\ \citenamefont {McMahon}}]{wright2022deep}%
  \BibitemOpen
  \bibfield  {author} {\bibinfo {author} {\bibfnamefont {Logan~G}\ \bibnamefont {Wright}}, \bibinfo {author} {\bibfnamefont {Tatsuhiro}\ \bibnamefont {Onodera}}, \bibinfo {author} {\bibfnamefont {Martin~M}\ \bibnamefont {Stein}}, \bibinfo {author} {\bibfnamefont {Tianyu}\ \bibnamefont {Wang}}, \bibinfo {author} {\bibfnamefont {Darren~T}\ \bibnamefont {Schachter}}, \bibinfo {author} {\bibfnamefont {Zoey}\ \bibnamefont {Hu}}, \ and\ \bibinfo {author} {\bibfnamefont {Peter~L}\ \bibnamefont {McMahon}},\ }\bibfield  {title} {\enquote {\bibinfo {title} {Deep physical neural networks trained with backpropagation},}\ }\href@noop {} {\bibfield  {journal} {\bibinfo  {journal} {Nature}\ }\textbf {\bibinfo {volume} {601}},\ \bibinfo {pages} {549--555} (\bibinfo {year} {2022})}\BibitemShut {NoStop}%
\bibitem [{\citenamefont {Guo}\ \emph {et~al.}(2023)\citenamefont {Guo}, \citenamefont {Gutierrez}, \citenamefont {Sekine}, \citenamefont {Gray}, \citenamefont {Williams}, \citenamefont {Ledezma}, \citenamefont {Costa}, \citenamefont {Roy}, \citenamefont {Zhou}, \citenamefont {Liu} \emph {et~al.}}]{guo2023ultrafast}%
  \BibitemOpen
  \bibfield  {author} {\bibinfo {author} {\bibfnamefont {Qiushi}\ \bibnamefont {Guo}}, \bibinfo {author} {\bibfnamefont {Benjamin~K}\ \bibnamefont {Gutierrez}}, \bibinfo {author} {\bibfnamefont {Ryoto}\ \bibnamefont {Sekine}}, \bibinfo {author} {\bibfnamefont {Robert~M}\ \bibnamefont {Gray}}, \bibinfo {author} {\bibfnamefont {James~A}\ \bibnamefont {Williams}}, \bibinfo {author} {\bibfnamefont {Luis}\ \bibnamefont {Ledezma}}, \bibinfo {author} {\bibfnamefont {Luis}\ \bibnamefont {Costa}}, \bibinfo {author} {\bibfnamefont {Arkadev}\ \bibnamefont {Roy}}, \bibinfo {author} {\bibfnamefont {Selina}\ \bibnamefont {Zhou}}, \bibinfo {author} {\bibfnamefont {Mingchen}\ \bibnamefont {Liu}},  \emph {et~al.},\ }\bibfield  {title} {\enquote {\bibinfo {title} {Ultrafast mode-locked laser in nanophotonic lithium niobate},}\ }\href@noop {} {\bibfield  {journal} {\bibinfo  {journal} {Science}\ }\textbf {\bibinfo {volume} {382}},\ \bibinfo {pages} {708--713} (\bibinfo {year} {2023})}\BibitemShut {NoStop}%
\bibitem [{\citenamefont {Bogaerts}\ \emph {et~al.}(2020)\citenamefont {Bogaerts}, \citenamefont {P{\'e}rez}, \citenamefont {Capmany}, \citenamefont {Miller}, \citenamefont {Poon}, \citenamefont {Englund}, \citenamefont {Morichetti},\ and\ \citenamefont {Melloni}}]{bogaerts2020programmable}%
  \BibitemOpen
  \bibfield  {author} {\bibinfo {author} {\bibfnamefont {Wim}\ \bibnamefont {Bogaerts}}, \bibinfo {author} {\bibfnamefont {Daniel}\ \bibnamefont {P{\'e}rez}}, \bibinfo {author} {\bibfnamefont {Jos{\'e}}\ \bibnamefont {Capmany}}, \bibinfo {author} {\bibfnamefont {David~AB}\ \bibnamefont {Miller}}, \bibinfo {author} {\bibfnamefont {Joyce}\ \bibnamefont {Poon}}, \bibinfo {author} {\bibfnamefont {Dirk}\ \bibnamefont {Englund}}, \bibinfo {author} {\bibfnamefont {Francesco}\ \bibnamefont {Morichetti}}, \ and\ \bibinfo {author} {\bibfnamefont {Andrea}\ \bibnamefont {Melloni}},\ }\bibfield  {title} {\enquote {\bibinfo {title} {Programmable photonic circuits},}\ }\href@noop {} {\bibfield  {journal} {\bibinfo  {journal} {Nature}\ }\textbf {\bibinfo {volume} {586}},\ \bibinfo {pages} {207--216} (\bibinfo {year} {2020})}\BibitemShut {NoStop}%
\bibitem [{\citenamefont {Shan}\ \emph {et~al.}(2021)\citenamefont {Shan}, \citenamefont {Ye}, \citenamefont {Chu}, \citenamefont {Lee}, \citenamefont {Park}, \citenamefont {Balents},\ and\ \citenamefont {Hsieh}}]{shan2021giant}%
  \BibitemOpen
  \bibfield  {author} {\bibinfo {author} {\bibfnamefont {Jun-Yi}\ \bibnamefont {Shan}}, \bibinfo {author} {\bibfnamefont {M}~\bibnamefont {Ye}}, \bibinfo {author} {\bibfnamefont {H}~\bibnamefont {Chu}}, \bibinfo {author} {\bibfnamefont {Sungmin}\ \bibnamefont {Lee}}, \bibinfo {author} {\bibfnamefont {Je-Geun}\ \bibnamefont {Park}}, \bibinfo {author} {\bibfnamefont {L}~\bibnamefont {Balents}}, \ and\ \bibinfo {author} {\bibfnamefont {D}~\bibnamefont {Hsieh}},\ }\bibfield  {title} {\enquote {\bibinfo {title} {Giant modulation of optical nonlinearity by {Floquet} engineering},}\ }\href@noop {} {\bibfield  {journal} {\bibinfo  {journal} {Nature}\ }\textbf {\bibinfo {volume} {600}},\ \bibinfo {pages} {235--239} (\bibinfo {year} {2021})}\BibitemShut {NoStop}%
\bibitem [{\citenamefont {Sie}\ \emph {et~al.}(2019)\citenamefont {Sie}, \citenamefont {Nyby}, \citenamefont {Pemmaraju}, \citenamefont {Park}, \citenamefont {Shen}, \citenamefont {Yang}, \citenamefont {Hoffmann}, \citenamefont {Ofori-Okai}, \citenamefont {Li}, \citenamefont {Reid} \emph {et~al.}}]{sie2019ultrafast}%
  \BibitemOpen
  \bibfield  {author} {\bibinfo {author} {\bibfnamefont {Edbert~J}\ \bibnamefont {Sie}}, \bibinfo {author} {\bibfnamefont {Clara~M}\ \bibnamefont {Nyby}}, \bibinfo {author} {\bibfnamefont {CD}~\bibnamefont {Pemmaraju}}, \bibinfo {author} {\bibfnamefont {Su~Ji}\ \bibnamefont {Park}}, \bibinfo {author} {\bibfnamefont {Xiaozhe}\ \bibnamefont {Shen}}, \bibinfo {author} {\bibfnamefont {Jie}\ \bibnamefont {Yang}}, \bibinfo {author} {\bibfnamefont {Matthias~C}\ \bibnamefont {Hoffmann}}, \bibinfo {author} {\bibfnamefont {BK}~\bibnamefont {Ofori-Okai}}, \bibinfo {author} {\bibfnamefont {Renkai}\ \bibnamefont {Li}}, \bibinfo {author} {\bibfnamefont {Alexander~H}\ \bibnamefont {Reid}},  \emph {et~al.},\ }\bibfield  {title} {\enquote {\bibinfo {title} {An ultrafast symmetry switch in a {Weyl} semimetal},}\ }\href@noop {} {\bibfield  {journal} {\bibinfo  {journal} {Nature}\ }\textbf {\bibinfo {volume} {565}},\ \bibinfo {pages} {61--66} (\bibinfo {year} {2019})}\BibitemShut {NoStop}%
\bibitem [{\citenamefont {de~la Torre}\ \emph {et~al.}(2022)\citenamefont {de~la Torre}, \citenamefont {Seyler}, \citenamefont {Buchhold}, \citenamefont {Baum}, \citenamefont {Zhang}, \citenamefont {Laurita}, \citenamefont {Harter}, \citenamefont {Zhao}, \citenamefont {Phinney}, \citenamefont {Chen} \emph {et~al.}}]{de2022decoupling}%
  \BibitemOpen
  \bibfield  {author} {\bibinfo {author} {\bibfnamefont {Alberto}\ \bibnamefont {de~la Torre}}, \bibinfo {author} {\bibfnamefont {Kyle~L}\ \bibnamefont {Seyler}}, \bibinfo {author} {\bibfnamefont {Michael}\ \bibnamefont {Buchhold}}, \bibinfo {author} {\bibfnamefont {Yuval}\ \bibnamefont {Baum}}, \bibinfo {author} {\bibfnamefont {Gufeng}\ \bibnamefont {Zhang}}, \bibinfo {author} {\bibfnamefont {Nicholas~J}\ \bibnamefont {Laurita}}, \bibinfo {author} {\bibfnamefont {John~W}\ \bibnamefont {Harter}}, \bibinfo {author} {\bibfnamefont {Liuyan}\ \bibnamefont {Zhao}}, \bibinfo {author} {\bibfnamefont {Isabelle}\ \bibnamefont {Phinney}}, \bibinfo {author} {\bibfnamefont {Xiang}\ \bibnamefont {Chen}},  \emph {et~al.},\ }\bibfield  {title} {\enquote {\bibinfo {title} {Decoupling of static and dynamic criticality in a driven {Mott} insulator},}\ }\href@noop {} {\bibfield  {journal} {\bibinfo  {journal} {Commun. Phys.}\ }\textbf {\bibinfo {volume} {5}},\ \bibinfo {pages} {35} (\bibinfo {year} {2022})}\BibitemShut
  {NoStop}%
\bibitem [{\citenamefont {Boyd}(2003)}]{boyd}%
  \BibitemOpen
  \bibfield  {author} {\bibinfo {author} {\bibfnamefont {R.~W.}\ \bibnamefont {Boyd}},\ }\href@noop {} {\emph {\bibinfo {title} {Nonlinear {{Optics}}}}}\ (\bibinfo  {publisher} {{Academic Press, Burlington, MA}},\ \bibinfo {year} {2003})\BibitemShut {NoStop}%
\bibitem [{\citenamefont {Chu}\ \emph {et~al.}(2020)\citenamefont {Chu}, \citenamefont {Roh}, \citenamefont {Island}, \citenamefont {Li}, \citenamefont {Lee}, \citenamefont {Chen}, \citenamefont {Park}, \citenamefont {Young}, \citenamefont {Lee},\ and\ \citenamefont {Hsieh}}]{chu2020linear}%
  \BibitemOpen
  \bibfield  {author} {\bibinfo {author} {\bibfnamefont {Hao}\ \bibnamefont {Chu}}, \bibinfo {author} {\bibfnamefont {Chang~Jae}\ \bibnamefont {Roh}}, \bibinfo {author} {\bibfnamefont {Joshua~O}\ \bibnamefont {Island}}, \bibinfo {author} {\bibfnamefont {Chen}\ \bibnamefont {Li}}, \bibinfo {author} {\bibfnamefont {Sungmin}\ \bibnamefont {Lee}}, \bibinfo {author} {\bibfnamefont {Jingjing}\ \bibnamefont {Chen}}, \bibinfo {author} {\bibfnamefont {Je-Geun}\ \bibnamefont {Park}}, \bibinfo {author} {\bibfnamefont {Andrea~F}\ \bibnamefont {Young}}, \bibinfo {author} {\bibfnamefont {Jong~Seok}\ \bibnamefont {Lee}}, \ and\ \bibinfo {author} {\bibfnamefont {David}\ \bibnamefont {Hsieh}},\ }\bibfield  {title} {\enquote {\bibinfo {title} {Linear magnetoelectric phase in ultrathin {MnPS$_3$} probed by optical second harmonic generation},}\ }\href@noop {} {\bibfield  {journal} {\bibinfo  {journal} {Phys. Rev. Lett.}\ }\textbf {\bibinfo {volume} {124}},\ \bibinfo {pages} {027601} (\bibinfo {year} {2020})}\BibitemShut
  {NoStop}%
\bibitem [{\citenamefont {Kang}\ \emph {et~al.}(2020)\citenamefont {Kang}, \citenamefont {Kim}, \citenamefont {Kim}, \citenamefont {Kim}, \citenamefont {Sim}, \citenamefont {Lee}, \citenamefont {Lee}, \citenamefont {Park}, \citenamefont {Yun}, \citenamefont {Kim} \emph {et~al.}}]{kang2020coherent}%
  \BibitemOpen
  \bibfield  {author} {\bibinfo {author} {\bibfnamefont {Soonmin}\ \bibnamefont {Kang}}, \bibinfo {author} {\bibfnamefont {Kangwon}\ \bibnamefont {Kim}}, \bibinfo {author} {\bibfnamefont {Beom~Hyun}\ \bibnamefont {Kim}}, \bibinfo {author} {\bibfnamefont {Jonghyeon}\ \bibnamefont {Kim}}, \bibinfo {author} {\bibfnamefont {Kyung~Ik}\ \bibnamefont {Sim}}, \bibinfo {author} {\bibfnamefont {Jae-Ung}\ \bibnamefont {Lee}}, \bibinfo {author} {\bibfnamefont {Sungmin}\ \bibnamefont {Lee}}, \bibinfo {author} {\bibfnamefont {Kisoo}\ \bibnamefont {Park}}, \bibinfo {author} {\bibfnamefont {Seokhwan}\ \bibnamefont {Yun}}, \bibinfo {author} {\bibfnamefont {Taehun}\ \bibnamefont {Kim}},  \emph {et~al.},\ }\bibfield  {title} {\enquote {\bibinfo {title} {Coherent many-body exciton in van der {Waals} antiferromagnet {NiPS$_3$}},}\ }\href@noop {} {\bibfield  {journal} {\bibinfo  {journal} {Nature}\ }\textbf {\bibinfo {volume} {583}},\ \bibinfo {pages} {785--789} (\bibinfo {year} {2020})}\BibitemShut {NoStop}%
\bibitem [{\citenamefont {Chen}\ \emph {et~al.}(2021)\citenamefont {Chen}, \citenamefont {Qi}, \citenamefont {Liu}, \citenamefont {Chen}, \citenamefont {Wang}, \citenamefont {Yan}, \citenamefont {Zhang}, \citenamefont {Cao}, \citenamefont {Lu}, \citenamefont {Tian} \emph {et~al.}}]{chen2021electrically}%
  \BibitemOpen
  \bibfield  {author} {\bibinfo {author} {\bibfnamefont {Guangyi}\ \bibnamefont {Chen}}, \bibinfo {author} {\bibfnamefont {Shaomian}\ \bibnamefont {Qi}}, \bibinfo {author} {\bibfnamefont {Jianqiao}\ \bibnamefont {Liu}}, \bibinfo {author} {\bibfnamefont {Di}~\bibnamefont {Chen}}, \bibinfo {author} {\bibfnamefont {Jiongjie}\ \bibnamefont {Wang}}, \bibinfo {author} {\bibfnamefont {Shili}\ \bibnamefont {Yan}}, \bibinfo {author} {\bibfnamefont {Yu}~\bibnamefont {Zhang}}, \bibinfo {author} {\bibfnamefont {Shimin}\ \bibnamefont {Cao}}, \bibinfo {author} {\bibfnamefont {Ming}\ \bibnamefont {Lu}}, \bibinfo {author} {\bibfnamefont {Shibing}\ \bibnamefont {Tian}},  \emph {et~al.},\ }\bibfield  {title} {\enquote {\bibinfo {title} {Electrically switchable van der {Waals} magnon valves},}\ }\href@noop {} {\bibfield  {journal} {\bibinfo  {journal} {Nat. Commun.}\ }\textbf {\bibinfo {volume} {12}},\ \bibinfo {pages} {6279} (\bibinfo {year} {2021})}\BibitemShut {NoStop}%
\bibitem [{\citenamefont {Belvin}\ \emph {et~al.}(2021)\citenamefont {Belvin}, \citenamefont {Baldini}, \citenamefont {Ozel}, \citenamefont {Mao}, \citenamefont {Po}, \citenamefont {Allington}, \citenamefont {Son}, \citenamefont {Kim}, \citenamefont {Kim}, \citenamefont {Hwang} \emph {et~al.}}]{belvin2021exciton}%
  \BibitemOpen
  \bibfield  {author} {\bibinfo {author} {\bibfnamefont {Carina~A}\ \bibnamefont {Belvin}}, \bibinfo {author} {\bibfnamefont {Edoardo}\ \bibnamefont {Baldini}}, \bibinfo {author} {\bibfnamefont {Ilkem~Ozge}\ \bibnamefont {Ozel}}, \bibinfo {author} {\bibfnamefont {Dan}\ \bibnamefont {Mao}}, \bibinfo {author} {\bibfnamefont {Hoi~Chun}\ \bibnamefont {Po}}, \bibinfo {author} {\bibfnamefont {Clifford~J}\ \bibnamefont {Allington}}, \bibinfo {author} {\bibfnamefont {Suhan}\ \bibnamefont {Son}}, \bibinfo {author} {\bibfnamefont {Beom~Hyun}\ \bibnamefont {Kim}}, \bibinfo {author} {\bibfnamefont {Jonghyeon}\ \bibnamefont {Kim}}, \bibinfo {author} {\bibfnamefont {Inho}\ \bibnamefont {Hwang}},  \emph {et~al.},\ }\bibfield  {title} {\enquote {\bibinfo {title} {Exciton-driven antiferromagnetic metal in a correlated van der {Waals} insulator},}\ }\href@noop {} {\bibfield  {journal} {\bibinfo  {journal} {Nat. Commun.}\ }\textbf {\bibinfo {volume} {12}},\ \bibinfo {pages} {4837} (\bibinfo {year} {2021})}\BibitemShut {NoStop}%
\bibitem [{\citenamefont {He}\ \emph {et~al.}(2024)\citenamefont {He}, \citenamefont {Shen}, \citenamefont {Wohlfeld}, \citenamefont {Sears}, \citenamefont {Li}, \citenamefont {Pelliciari}, \citenamefont {Walicki}, \citenamefont {Johnston}, \citenamefont {Baldini}, \citenamefont {Bisogni} \emph {et~al.}}]{he2024magnetically}%
  \BibitemOpen
  \bibfield  {author} {\bibinfo {author} {\bibfnamefont {Wei}\ \bibnamefont {He}}, \bibinfo {author} {\bibfnamefont {Yao}\ \bibnamefont {Shen}}, \bibinfo {author} {\bibfnamefont {Krzysztof}\ \bibnamefont {Wohlfeld}}, \bibinfo {author} {\bibfnamefont {Jennifer}\ \bibnamefont {Sears}}, \bibinfo {author} {\bibfnamefont {Jiemin}\ \bibnamefont {Li}}, \bibinfo {author} {\bibfnamefont {Jonathan}\ \bibnamefont {Pelliciari}}, \bibinfo {author} {\bibfnamefont {M}~\bibnamefont {Walicki}}, \bibinfo {author} {\bibfnamefont {S}~\bibnamefont {Johnston}}, \bibinfo {author} {\bibfnamefont {E}~\bibnamefont {Baldini}}, \bibinfo {author} {\bibfnamefont {V}~\bibnamefont {Bisogni}},  \emph {et~al.},\ }\bibfield  {title} {\enquote {\bibinfo {title} {Magnetically propagating hund’s exciton in van der {Waals} antiferromagnet {NiPS$_3$}},}\ }\href@noop {} {\bibfield  {journal} {\bibinfo  {journal} {Nat. Commun.}\ }\textbf {\bibinfo {volume} {15}},\ \bibinfo {pages} {3496} (\bibinfo {year} {2024})}\BibitemShut {NoStop}%
\bibitem [{\citenamefont {Kim}\ \emph {et~al.}(2018)\citenamefont {Kim}, \citenamefont {Kim}, \citenamefont {Sandilands}, \citenamefont {Sinn}, \citenamefont {Lee}, \citenamefont {Son}, \citenamefont {Lee}, \citenamefont {Choi}, \citenamefont {Kim}, \citenamefont {Park} \emph {et~al.}}]{kim2018charge}%
  \BibitemOpen
  \bibfield  {author} {\bibinfo {author} {\bibfnamefont {So~Yeun}\ \bibnamefont {Kim}}, \bibinfo {author} {\bibfnamefont {Tae~Yun}\ \bibnamefont {Kim}}, \bibinfo {author} {\bibfnamefont {Luke~J}\ \bibnamefont {Sandilands}}, \bibinfo {author} {\bibfnamefont {Soobin}\ \bibnamefont {Sinn}}, \bibinfo {author} {\bibfnamefont {Min-Cheol}\ \bibnamefont {Lee}}, \bibinfo {author} {\bibfnamefont {Jaeseok}\ \bibnamefont {Son}}, \bibinfo {author} {\bibfnamefont {Sungmin}\ \bibnamefont {Lee}}, \bibinfo {author} {\bibfnamefont {Ki-Young}\ \bibnamefont {Choi}}, \bibinfo {author} {\bibfnamefont {Wondong}\ \bibnamefont {Kim}}, \bibinfo {author} {\bibfnamefont {Byeong-Gyu}\ \bibnamefont {Park}},  \emph {et~al.},\ }\bibfield  {title} {\enquote {\bibinfo {title} {Charge-spin correlation in van der {Waals} antiferromagnet {NiPS$_3$}},}\ }\href@noop {} {\bibfield  {journal} {\bibinfo  {journal} {Phys. Rev. Lett.}\ }\textbf {\bibinfo {volume} {120}},\ \bibinfo {pages} {136402} (\bibinfo {year} {2018})}\BibitemShut {NoStop}%
\bibitem [{\citenamefont {Harter}\ \emph {et~al.}(2015)\citenamefont {Harter}, \citenamefont {Niu}, \citenamefont {Woss},\ and\ \citenamefont {Hsieh}}]{harter2015high}%
  \BibitemOpen
  \bibfield  {author} {\bibinfo {author} {\bibfnamefont {JW}~\bibnamefont {Harter}}, \bibinfo {author} {\bibfnamefont {L}~\bibnamefont {Niu}}, \bibinfo {author} {\bibfnamefont {AJ}~\bibnamefont {Woss}}, \ and\ \bibinfo {author} {\bibfnamefont {David}\ \bibnamefont {Hsieh}},\ }\bibfield  {title} {\enquote {\bibinfo {title} {High-speed measurement of rotational anisotropy nonlinear optical harmonic generation using position-sensitive detection},}\ }\href@noop {} {\bibfield  {journal} {\bibinfo  {journal} {Opt. Lett.}\ }\textbf {\bibinfo {volume} {40}},\ \bibinfo {pages} {4671--4674} (\bibinfo {year} {2015})}\BibitemShut {NoStop}%
\bibitem [{\citenamefont {Lane}\ and\ \citenamefont {Zhu}(2020)}]{lane2020thickness}%
  \BibitemOpen
  \bibfield  {author} {\bibinfo {author} {\bibfnamefont {Christopher}\ \bibnamefont {Lane}}\ and\ \bibinfo {author} {\bibfnamefont {Jian-Xin}\ \bibnamefont {Zhu}},\ }\bibfield  {title} {\enquote {\bibinfo {title} {Thickness dependence of electronic structure and optical properties of a correlated van der {Waals} antiferromagnetic {NiPS$_3$} thin film},}\ }\href@noop {} {\bibfield  {journal} {\bibinfo  {journal} {Phys. Rev. B}\ }\textbf {\bibinfo {volume} {102}},\ \bibinfo {pages} {075124} (\bibinfo {year} {2020})}\BibitemShut {NoStop}%
\bibitem [{\citenamefont {Kirilyuk}\ \emph {et~al.}(2010)\citenamefont {Kirilyuk}, \citenamefont {Kimel},\ and\ \citenamefont {Rasing}}]{kirilyuk2010ultrafast}%
  \BibitemOpen
  \bibfield  {author} {\bibinfo {author} {\bibfnamefont {Andrei}\ \bibnamefont {Kirilyuk}}, \bibinfo {author} {\bibfnamefont {Alexey~V}\ \bibnamefont {Kimel}}, \ and\ \bibinfo {author} {\bibfnamefont {Theo}\ \bibnamefont {Rasing}},\ }\bibfield  {title} {\enquote {\bibinfo {title} {Ultrafast optical manipulation of magnetic order},}\ }\href@noop {} {\bibfield  {journal} {\bibinfo  {journal} {Rev. Mod. Phys.}\ }\textbf {\bibinfo {volume} {82}},\ \bibinfo {pages} {2731--2784} (\bibinfo {year} {2010})}\BibitemShut {NoStop}%
\bibitem [{\citenamefont {Ho}\ \emph {et~al.}(2021)\citenamefont {Ho}, \citenamefont {Hsu},\ and\ \citenamefont {Muhimmah}}]{ho2021band}%
  \BibitemOpen
  \bibfield  {author} {\bibinfo {author} {\bibfnamefont {Ching-Hwa}\ \bibnamefont {Ho}}, \bibinfo {author} {\bibfnamefont {Tien-Yao}\ \bibnamefont {Hsu}}, \ and\ \bibinfo {author} {\bibfnamefont {Luthviyah~Choirotul}\ \bibnamefont {Muhimmah}},\ }\bibfield  {title} {\enquote {\bibinfo {title} {The band-edge excitons observed in few-layer {NiPS$_3$}},}\ }\href@noop {} {\bibfield  {journal} {\bibinfo  {journal} {npj 2D Mater. Appl.}\ }\textbf {\bibinfo {volume} {5}},\ \bibinfo {pages} {8} (\bibinfo {year} {2021})}\BibitemShut {NoStop}%
\bibitem [{\citenamefont {Erge{\c{c}}en}\ \emph {et~al.}(2022)\citenamefont {Erge{\c{c}}en}, \citenamefont {Ilyas}, \citenamefont {Mao}, \citenamefont {Po}, \citenamefont {Yilmaz}, \citenamefont {Kim}, \citenamefont {Park}, \citenamefont {Senthil},\ and\ \citenamefont {Gedik}}]{ergeccen2022magnetically}%
  \BibitemOpen
  \bibfield  {author} {\bibinfo {author} {\bibfnamefont {Emre}\ \bibnamefont {Erge{\c{c}}en}}, \bibinfo {author} {\bibfnamefont {Batyr}\ \bibnamefont {Ilyas}}, \bibinfo {author} {\bibfnamefont {Dan}\ \bibnamefont {Mao}}, \bibinfo {author} {\bibfnamefont {Hoi~Chun}\ \bibnamefont {Po}}, \bibinfo {author} {\bibfnamefont {Mehmet~Burak}\ \bibnamefont {Yilmaz}}, \bibinfo {author} {\bibfnamefont {Junghyun}\ \bibnamefont {Kim}}, \bibinfo {author} {\bibfnamefont {Je-Geun}\ \bibnamefont {Park}}, \bibinfo {author} {\bibfnamefont {T}~\bibnamefont {Senthil}}, \ and\ \bibinfo {author} {\bibfnamefont {Nuh}\ \bibnamefont {Gedik}},\ }\bibfield  {title} {\enquote {\bibinfo {title} {Magnetically brightened dark electron-phonon bound states in a van der waals antiferromagnet},}\ }\href@noop {} {\bibfield  {journal} {\bibinfo  {journal} {Nat. Commun.}\ }\textbf {\bibinfo {volume} {13}},\ \bibinfo {pages} {98} (\bibinfo {year} {2022})}\BibitemShut {NoStop}%
\end{thebibliography}%
\end{document}